\documentclass[pra, twocolumn, amsmath,amssymb, aps]{revtex4-2}
 \usepackage{dcolumn}
\usepackage{bm}
\usepackage{graphicx}
\usepackage[caption=false]{subfig}
\usepackage{floatrow}
\usepackage{comment}
\usepackage{amsmath}
\usepackage{geometry}
\usepackage{mathrsfs}
\usepackage{algorithmic}
\usepackage{tensor}
\usepackage{physics}
\usepackage{float}
\usepackage{hyperref}

\begin{document}

\preprint{APS/123-QED}

\title{Dark matter condensates as highly nonlocal solitons: instability in the Schwarzschild metric and laboratory analog}

\author{Ludovica Dieli}
 \email{ludovica.dieli@uniroma1.it}
\author{Claudio Conti}%
\email{claudio.conti@uniroma1.it}
\affiliation{Dipartimento di Fisica, Università di Roma "La Sapienza", 00185 Rome, Italy}%

\date{\today}

\begin{abstract}
Theories on the bosonic nature of dark matter are a promising alternative to the cold dark matter model. Here we consider a dark matter halo in the state of a Bose-Einstein condensate, subject to the gravitation of a black hole. In the low energy limit, we bring together the general relativity in the Schwarzschild metric and the quantum description of the Bose-Einstein condensate. The model is solvable in the Fermi normal coordinates with the so called highly nonlocal approximation and describes tidal deformations in the condensate wave function. The black hole deforms the localized condensate until the attraction of the compact object overcomes the self-gravitation and destabilizes the solitonic dark matter. 
Moreover, the model can be implemented as a gravitational analog in the laboratory; the time-dependent potential generated by the galactic black hole can be mimicked by an optical trap acting on a conventional condensate.  The results open the way to new laboratory simulators for quantum gravitational effects.
\end{abstract}

\maketitle

\section{Introduction}
One of the main unsolved questions about the Universe is the nature of dark matter (DM). Many astrophysical observations have revealed the presence of a new kind of matter, which gravitationally interacts with the visible matter but does not emit light \cite{DM_review}. Among the proposed theories, one of the most promising is the cold dark matter model (CDM), based on collisionless particles. Despite CDM explains anomalous galaxy rotation curves, new discrepancies, such as the cuspy problem regarding the flat density distribution in the galactic core, remain not yet understood~\cite{DM_review}. Authors proposed the introduction of particle self-interaction to solve the inconsistencies in the CDM~\cite{DM_BEC_review}. Inspired by early Universe theories, bosonic DM with ultrasmall masses $\sim 10^{-24}-10^{-20}\, \text{eV}$ ~\cite{1994} allows the formation of galactic Bose-Einstein condensates (BECs)~\cite{bohmer2007, harko2020, cosmicstructure, paredes2016interference, garnier2021incoherent}. One key aspect of BEC DM halos is the role of spacetime curvature first considered in~\cite{bohmer2007}. In~\cite{DM_scalar_field}, the curved spacetime appears as a perturbative potential inducing an instability dependent on the condensate mass. However, the effect of an external gravitational field on the dynamics of the galactic condensate is unexplored.\\ \\ 
In this manuscript, we study a DM condensate subject to a rotationally symmetric gravitational field described by the Schwarzschild metric. 
The Bose-condensed DM experiences the gravitational field of a black hole (BH) in the form of an effective potential in the non-relativistic limit. By a simple closed-form theory, we show how the interaction induces the excitation of high-energy states.
These states are required to fit the galaxy rotation curves in~\cite{1994}. Also, we predict the existence of a breaking time for the destabilization of the BEC, which may explain the absence of DM in specific galaxies.\\
Moreover, we consider the feasibility of BEC DM simulations in a laboratory~\cite{BEC_EM_int}. We detail an experimental setup to address the gravitational instability in an optically trapped Bose gas. Indeed, in the context of analog gravity~\cite{barcelo2011analogue}, physical systems such as in polaritonics, atomic gases, and nonlinear optics are used to mimic the cosmological models.\\ 
The manuscript is organized as follows. In Sec.~\ref{the_model}, we introduce the model: $(i)$ the Fermi normal coordinates used to make the expansion of the d'Alembert operator and $(ii)$ the multiple scale expansion of the solution of the Klein-Gordon equation in the non-relativistic limit.\\
Section~\ref{pg:metric} is about the effective potential in the Schwarzschild metric, simplified by the highly-nonlocal approximation in Sec.~\ref{ASA}.\\
In Sec.~\ref{eq_study}, we analyze the asymptotic model and predict the existence of a critical time for the instability dependent on the masses of the DM and of the BH.\\ In Sec.~\ref{astro_exp}, we specifically address the significant timescales and other parameters in the interaction with BHs; in Sec.~\ref{lab_exp}, we described an experiment for the laboratory analog.
Conclusions are drawn in Sec.~\ref{conclusions}.  
\section{Model and methods}\label{the_model}
The hypothesis of BEC dark matter halos opens the question about the effects of the spacetime curvature. We analyze the scenario in which a BEC DM is freely falling in a curved metric. We take into account two aspects: $(i)$ the BEC with its quantum description and $(ii)$ the relativistic spacetime.\\
Concerning point $(i)$, we describe the evolution of the self-gravitating BEC by the Gross-Pitaevskii equation (GPE) \cite{dalfovo1999theoryBEC} with a gravitational potential
\begin{gather}
   i\hbar \partial_t \psi(t,\vec{r})=-\frac{\hbar^2}{2m} \nabla^2 \psi(t,\vec{r})+m\Phi(t,\vec{r}) \psi(t,\vec{r});\\
     \nabla^2 \Phi (t,\vec{r})-K^2 \Phi (t,\vec{r})= 4 \pi G m |\psi (t,\vec{r})|^2;
    \label{NSE}
\end{gather}
where $K$ is inversely proportional to the interaction length of the potential $\Phi$ and $G$ is the universal gravitational constant. $\Phi(t,\Vec{r})$ is a nonlinear long-range gravitational potential solution of the Poisson equation with the mass density distribution $m|\psi(t,\vec{r})|^2$. The BEC description is clearly non-relativistic since the Poisson equation concerns Newtonian gravitational field. However, eq.(\ref{NSE}) is naturally generalised to a relativistic scenario in the form of Einstein's equations \cite{gravitation}. \\ Concerning point $(ii)$, the BEC experiences an external gravitational field described by a metric $g_{\mu\nu}$. However, the GPE is valid in a non-relativistic regime, while the curved metric requires a four-dimensional formalism.\\
To unify the two descriptions in a non-relativistic regime, we expand the d'Alembert operator by means of the Fermi normal coordinates suitable to describe a freely falling object. The expansion is applied to the Klein-Gordon equation recovering a Schroedinger-like equation in the non-relativistic limit~\cite{exirifard2022schrodinger}. We also use a multiple scale expansion and treat the nonlinearity as a mass-dependent potential within the highly nonlocal approximation.
\subsection{Curved metric in Fermi normal coordinates}
\begin{figure} 
    \centering
    \includegraphics[width=\linewidth]{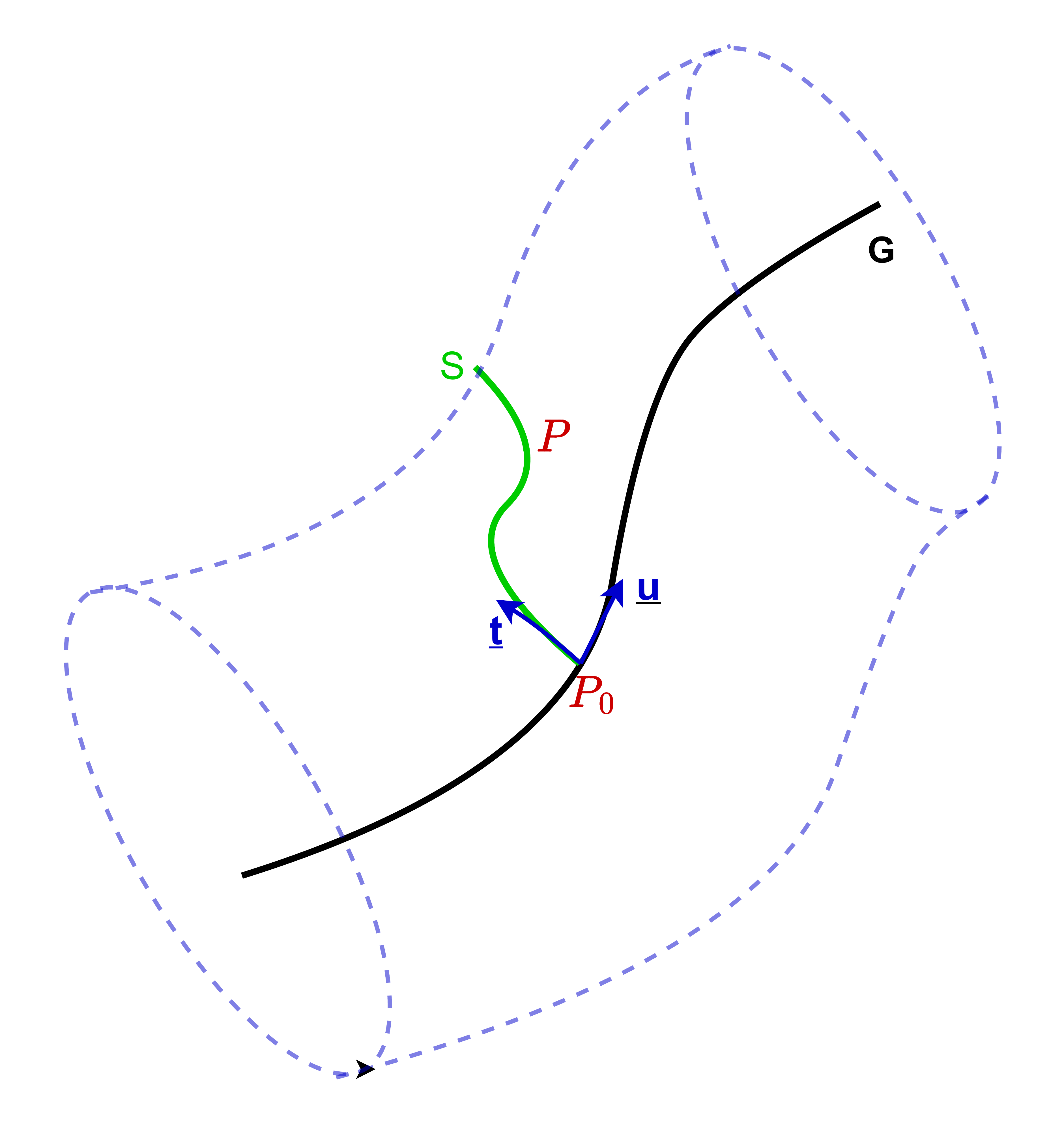}
    \caption{The timelike geodesic $G$ (black line) is surrounded by a world tube in which Fermi normal coordinates are defined, and special relativity laws are valid. The green line represents a spacelike geodesic $S$ on an orthogonal hypersurface $\Sigma$. The time component is tangent to the geodesic $G$: $\boldsymbol{e}_0=\boldsymbol{u}$, while the space components lie on the surface $\Sigma$: $\boldsymbol{t}=t^j \boldsymbol{e}_j$, $j=1,2,3$. In the world tube, the metric is flat up to quadratic corrections.}
    \label{fig:worldtube}
\end{figure}

According to the Equivalence Principle, a freely falling object is locally described by the flat metric. The Fermi normal coordinates enlarge this description to a world tube around the timelike trajectory~\cite{misner}. The Fermi normal coordinates satisfy the condition $\Gamma^\alpha_{\mu\nu}=0$ along the object trajectory. The basis vectors are defined in the following way: the time coordinate $\tau$ is such that $\boldsymbol{e_0}=\frac{\partial }{\partial \tau}$ is tangent to the trajectory. From every point of the trajectory, an orthogonal hypersurface $\Sigma$ is defined, and $3$ independent vectors are chosen on $\Sigma$ as the spatial vectors of the basis $\boldsymbol{e_j}=\frac{\partial}{\partial x_j}$ with $j=1,2,3$. A $(2+1)D$ simplified representation is reported in Fig.~\ref{fig:worldtube}. $(c\tau, x_1, x_2, x_3)$ are the Fermi normal coordinates. They describe the spacetime around the falling object as an expansion of the flat metric
\begin{equation*}
\begin{split}
    g_{\mu\nu}(c\tau, \Vec{x})&=\eta_{\mu\nu}+g_{\mu\nu, \alpha}(c\tau,0) x^{\alpha}\\
    &+\frac{1}{2} g_{\mu\nu,\alpha\beta}(c\tau,0) x^\alpha x^\beta + o(x^2)
    \end{split}
\end{equation*}
where the comma stands for the partial derivative of the metric tensor with respect to the Fermi normal coordinates.\\
The linear term vanishes due to $\Gamma^\alpha_{\mu\nu}=0$ on the timelike geodesic, and the first corrections are quadratic. According to~\cite{misner}, we found the following expressions for the metric elements
\begin{equation}
\begin{split}
&g_{00}=-1-R_{0l0m} x^lx^m + o(x^2); \\
&g_{0k}=-\frac{2}{3}R_{0jki} x^ix^j +o(x^2);\\
&g_{lm}=\delta_{lm}- \frac{1}{3}R_{iljm} x^ix^j+o(x^2);
\end{split}
\label{metric_elements}
\end{equation}
where $R_{\alpha \beta \mu\nu}$ is the Riemann tensor. The quadratic terms are small in the neighborhood of the geodesic, defining $\varepsilon\ll 1$: $g_{\mu\nu}=\eta_{\mu\nu}+\varepsilon^2g^{(1)}_{\mu\nu}$. Thus, given the spacetime metric $g_{\mu\nu}$, its expansion is defined by the Riemann tensor in Fermi normal coordinates.

We use this metric expansion to calculate the d'Alembert operator in the curved spacetime as a series expansion defined in the neighbor of the geodesic.\\ The d'Alembertian in general coordinates is
\begin{equation*}
\square=\frac{1}{\sqrt{g}}\partial_\mu(\sqrt{g} g^{\mu\nu}\partial_\nu)
\end{equation*}
where $g=-\text{det}(g_{\mu\nu})$, after~(\ref{metric_elements})
\begin{equation}
\begin{split}
 g&=-\det(g_{\mu \nu})=1+\varepsilon^2\frac{2}{3}R_{0l0m}\,x^lx^m+o(\varepsilon^2)\\
 &=1+\varepsilon^2 g^{(1)}+o(\varepsilon^2).
\end{split}
\label{determinant}
\end{equation}
The inverse matrix for the metric can be found from $g^{\mu\nu}g_{\nu\xi}=\delta^{\mu}_{\xi}$
\begin{equation*}
\begin{split}
&g^{\mu\nu}=\eta^{\mu\nu}+\varepsilon^2 g^{(1)\mu\nu}; \\
&g_{\nu\xi}=\eta_{\nu\xi}+\varepsilon^2 g^{(1)}_{\nu\xi}.
\end{split}
\label{inverse_metric}
\end{equation*}
We have
\begin{equation}
\begin{split}
&g^{(1)00}=-g^{(1)}_{00}=R_{0l0m}\,x^lx^m;\\
&g^{(1)0i}=-g^{(1)}_{0i}=\frac{2}{3}R_{0lim}\,x^lx^m;\\
&g^{(1)ij}=-g^{(1)}_{ij}=\frac{1}{3}R_{iljm}\,x^lx^m.
\label{g_espansion_elements}
\end{split}
\end{equation}\\
\begin{widetext}  
and
\begin{equation}
\begin{split}
\square&=\left(1-\frac{\varepsilon^2}{2}g^{(1)}\right)\,\partial_{\mu}\left[\left(1+\frac{\varepsilon^2}{2}g^{(1)}\right) \left(\eta^{\mu\nu}+\varepsilon^2 g^{(1)\mu\nu}\right)\partial_{\nu}\right]=\\
&=-\partial^2_{0}+\nabla^2+\varepsilon^2\partial_{\mu}(g^{(1)\mu\nu}\partial_{\nu})+\frac{\varepsilon^2}{2}\eta^{\mu\nu}\partial_{\mu}(g^{(1)})\partial_{\nu}+o(\varepsilon^2).
\end{split}
\label{exp_dalembert}
\end{equation}
\end{widetext}
This expansion in the Fermi normal coordinates needs a specific choice of the metric describing the falling object.

\subsection{Multiple scale expansion}
We start with the Klein-Gordon (KG) equation
\begin{equation}
    (\square -m^2)\phi=0,
    \label{KG}
\end{equation}
which reduces to the Schroedinger equation in the non-relativistic limit, and use Eq.~(\ref{exp_dalembert})
\begin{equation*}
\begin{split}
\square&=\square_{\text{flat}}+\varepsilon^2\partial_{\mu}(g^{(1)\mu\nu}\partial_{\nu})\\
    &+\frac{\varepsilon^2}{2}\eta^{\mu\nu} (\partial_{\mu}g^{(1)})\partial_\nu +o(\varepsilon^2).
\end{split}
\end{equation*}
Where we adopted units with $c=1$ and $\hbar=1$.\\
Since the effect of the curvature appears as a second-order perturbation of the flat d'Alembertian, we add the nonlinear potential of the DM condensate as a term of the same order
\begin{equation}
      (\square -m^2)\phi+\varepsilon^2 \Phi[\phi] \phi=0.
      \label{KG_phi}
\end{equation}
We define a multiple scale expansion by the slow variables $x_1^{\mu}=\varepsilon x^{\mu}$ and $x_2^{\mu}=\varepsilon^2 x^{\mu}$
\begin{equation}
\begin{split}
     \phi(x^{\mu}, x_1^{\mu}, x_2^{\mu})&=\phi_0(x^{\mu}, x_1^{\mu}, x_2^{\mu}) + \varepsilon \phi_1(x^{\mu}, x_1^{\mu}, x_2^{\mu}) \\
     &+\varepsilon^2 \phi_2(x^{\mu}, x_1^{\mu}, x_2^{\mu})+ o(\varepsilon^2).
  \end{split}  
\end{equation}
Due to the presence of the slow variables, the flat d'Alembertian is
\begin{equation*}
\begin{split}
\square_{\text{flat}}&=\square_0+2\varepsilon \square_{01}+\varepsilon^2 \square_1+2\varepsilon^2 \square_{02} + o(\varepsilon^2)\\
   & = \eta^{\mu\nu}\frac{\partial^2}{\partial x^{\mu}\partial x^{\nu}}+2\varepsilon \eta^{\mu\nu}\frac{\partial^2}{\partial x^{\mu}\partial x_1^{\nu}}\\
   &+\varepsilon^2  \eta^{\mu\nu} \frac{\partial^2}{\partial x_1^{\mu} \partial x_1^{\nu}}+2\varepsilon^2 \eta^{\mu\nu}\frac{\partial^2}{\partial x^{\mu} \partial x_2^{\nu}}+o(\varepsilon^2).
  \end{split}  
\end{equation*}
For the self-gravitating potential we have
\begin{equation}
\begin{split}
    \Phi[\phi]&=\int U(\vec{r}-\vec{r}\,') |\phi(\vec{r}\,')|^2 d^3r\,'\\
    &=\int U(\vec{r}-\vec{r}\,') [|\phi_0(\vec{r}\,')|^2+\varepsilon 2 \phi_0\phi_1\\
    &+\varepsilon^2|\phi_1(\vec{r}\,')|^2+\varepsilon^2 2 \phi_0\phi_2\\
    &+\varepsilon^3 2 \phi_1\phi_2+\varepsilon^4|\phi_2(\vec{r}\,')|^2] d^3r\,'.
    \end{split}
\end{equation}\\
\begin{widetext}
Equation~(\ref{KG_phi}) is
\begin{equation}
\begin{split} 
&\left[\square_0+2\varepsilon\square_{01}+\varepsilon^2 \square_1+2\varepsilon^2 \square_{02}+\varepsilon^2\partial_\mu (g^{(1)\mu\nu} \partial_\nu)+\varepsilon^2\frac{\eta^{\mu\nu}}{2}(\partial_\mu g^{(1)})\partial_\nu \right](\phi_0+\varepsilon\phi_1+\varepsilon^2 \phi_2)\\
    &-m^2(\phi_0+\varepsilon\phi_1+\varepsilon^2 \phi_2)+\varepsilon^2 \Phi[|\phi_0|^2] (\phi_0+\varepsilon\phi_1+\varepsilon^2 \phi_2)=0.
\end{split}
\label{eq:orders}
\end{equation}
\end{widetext}
At order $O(1)$ we have
\begin{equation}
  (\square_0-m^2)\phi_0=0   
\label{eq_ord_0}
\end{equation}
which is the KG equation in the flat spacetime. The solution is a free particle wave function modulated by an amplitude depending on the slow variables
\begin{equation*}
    \begin{split}
    \phi_0(x^{\mu}, x_1^{\mu}, x_2^{\mu})&=e^{ik_{\mu} x^{\mu}}\psi(x_1^{\mu}, x_2^{\mu})\\
     &+e^{-ik_{\mu} x^{\mu}}\psi^*(x_1^{\mu}, x_2^{\mu}).
     \end{split}
\end{equation*}
At the order $O(\varepsilon)$ in Eq.~(\ref{eq:orders}) we have
\begin{equation}
 (\square_0-m^2)\phi_1+2\square_{01} \phi_0=0   
\end{equation}
that is
\begin{multline}   
 (\square_0-m^2)\phi_1=-2\square_{01} \phi_0 \\
 =-2\eta^{\mu\nu}\frac{\partial^2 \phi_0}{\partial x^{\mu}\partial x_1^{\nu}}\\
 =-2i\omega e^{-i\omega \tau +i \vec{k}\cdot \vec{x}}\left(-\frac{\partial \psi}{\partial \tau_1}+\frac{k^j}{\omega}\frac{\partial \psi}{\partial x_1^j}\right)+c.c. 
  \label{Oesp1}
\end{multline}
We notice that the forcing term is resonant with the solution of the homogeneous equation: it has to be null to avoid secular terms
\begin{equation}
    \left(-\frac{\partial \psi}{\partial \tau_1}+\frac{k^j}{\omega}\frac{\partial \psi}{\partial x_1^j}\right) =0.
\label{constraint}
\end{equation}
As a consequence, Eq.~(\ref{Oesp1}) reduces to the KG equation, and we choose $\phi_1=0$.\\
After Eq.~(\ref{constraint}) we have
\begin{multline}   
\psi(x_1^{\mu}, x_2^{\mu})=\psi(\xi_1^j, x_2^{\mu});\\
\xi_1^j=k^j\left(x_1^j+\frac{k^j}{\omega}\tau_1\right) \quad j=1,2,3;
\end{multline}
with $k^j$ arbitrary functions. The upper index $j$ identifies the component of a $3$-dimensional vector.\\
At the order $O(\varepsilon^2)$ in Eq.~(\ref{eq:orders}) we find
\begin{equation}
    \begin{split}
        & \left[\square_1+2\square_{02}+\partial_\mu (g^{(1)\mu\nu}\partial_\nu)+\frac{\eta^{\mu\nu}}{2}(\partial_\mu g^{(1)})\partial_\nu \right]\phi_0\\
        &+ m\Phi[\phi_0]\phi_0+(\square_0-m^2)\phi_2=0.
    \end{split}
    \label{Oeps2}
\end{equation}
The gravitational potential is a function of $\phi_0$, thus
\begin{equation}
\begin{split}
\Phi[\phi_0]&=\int G(\vec{r}-\vec{r}\,')\left[2|\psi(\Vec{r}\,')|^2+\right.\\
    &\left.+2 \Re\left(e^{2ik_\mu x^\mu}\psi^2(\vec{r}\,')\right)\right]d^3r^\prime
    \end{split}
\end{equation}
this means that, solving the equation (\ref{Oeps2}) with respect to $\phi_2$, it needs the resonant term to vanish, obtaining the following 
\begin{equation}
\begin{split}
    &\left[\square_1+2\square_{02}+\partial_\mu (g^{(1)\mu\nu}\partial_\nu)+\frac{\eta^{\mu\nu}}{2}(\partial_\mu g^{(1)})\partial_\nu \right.\\
   &+ 2m\left.\int d^3r^\prime \, G(\vec{r}-\vec{r}\,')|\psi(\vec{r}\,')|^2\right]\phi_0(x^\mu, x_1^\mu, x_2^\mu); 
    \end{split}
    \label{eq_oeps2_1}
\end{equation}
and
\begin{multline}
(\square_0-m^2)\phi_2\\
+2m\phi_0\int d^3r^\prime G(\vec{r}-\vec{r}\,')\,\Re\left(e^{2ik_\mu x^\mu}\psi^2\right)=0.
\end{multline}
We consider the non-relativistic limit of the leading Eq.~(\ref{eq_oeps2_1}) by assuming a low energy regime with $\omega^2=m^2+k^2\simeq m^2$
\begin{equation}
    \begin{split}
       & \phi_0(x^\mu,x_1^\mu,x_2^\mu)=e^{-im\tau} \psi(\xi_1^j, x_2^\mu)+c.c.;\\
        &|\vec{k}\cdot \Vec{x}|\ll 1 , \quad \omega^2 \sim m^2.
    \end{split}
\end{equation}
As a consequence, the third and fourth terms of equation (\ref{eq_oeps2_1}) contain only time derivatives
\begin{equation*}
    \begin{split}
      &\partial_\mu (g^{(1)\mu\nu}\partial_\nu \phi_0)=(\partial_0 g^{(1)00}) \partial_0 \phi_0+g^{(1)00} \partial_0^2\phi_0\\
        &=[-im(\partial_0 g^{(1)00})-m^2g^{(1)00}] e^{-im\tau }\psi;
    \end{split}
\end{equation*}
and
\begin{equation*}
    \begin{split}        
    &\frac{\eta^{\mu\nu}}{2}(\partial_\mu g^{(1)})\partial_\nu \phi_0 =-\frac{1}{2}(\partial_0 g^{(1)})\partial_0 \phi_0\\
    &+\frac{1}{2}\delta^{kj}(\partial_k g^{(1)}) \partial_j \phi_0\\
    &=\frac{im}{2}(\partial_0 g^{(1)}) e^{-im\tau} \psi.
    \end{split}
\end{equation*}
From  Eq.~(\ref{g_espansion_elements}), $g^{(1)}$ and $g^{(1)00}$ depend on time $\tau$ only through the Riemann tensor, which is a function of the coordinate system centered on the BH. Noting that the time variation of the distance between the BEC and the BH is much smaller than the distance itself, namely $\partial_0g^{(1)00}\ll g^{(1)00}$ and $\partial_0g^{(1)}\ll g^{(1)00}$ we neglect the corresponding terms (as detailed for Schwarzschild metric in Sec.~\ref{pg:metric}).\\ 
The first two terms of the equation~(\ref{eq_oeps2_1}) are
\begin{equation*}
    \begin{split}
        &(\square_1+2\square_{02})\phi_0\\
        &=\left(\eta^{\mu\nu}\frac{\partial^2}{\partial x_1^\mu \partial x_1^\nu}+2\eta^{\mu\nu}\frac{\partial^2}{\partial x^\mu \partial x_2^\nu}\right) e^{-im\tau} \psi\\
        &=-e^{-im\tau} \frac{\partial^2 \psi}{\partial \tau_1^2}+e^{-im\tau}\nabla_1^2\psi+2im e^{-im\tau}\frac{\partial \psi}{\partial \tau_2}.
    \end{split}
\end{equation*}
Due to the constraint imposed on the slow time derivative, Eq.~(\ref{constraint}), we have
\begin{equation*}
    \begin{split}
        \frac{\partial^2\psi}{\partial \tau_1^2}&=\frac{\partial}{\partial \tau_1}\left(\frac{k^j}{\omega}\frac{\partial \psi}{\partial x_1^j}\right)\\
        &=\frac{\partial}{\partial \tau_1}\frac{k^2}{\omega}\frac{\partial \psi}{\partial \xi_1}
    \end{split}
\end{equation*}
and we can neglect this term due to the approximation $\omega^2\simeq m^2$ and $|k|^2\sim 0$.\\ Concluding the expansion by posing $\varepsilon=1$ and reinserting all the natural constants, the final equation results to be
\begin{equation}
i\hbar\frac{\partial \psi}{\partial \tau}=-\frac{\hbar^2}{2m} \nabla^2 \psi+\frac{mc^2}{2} g^{(1)00}\psi+m\Phi[|\psi|^2] \psi. 
    \label{general_model_eq}
 \end{equation}
 We highlight that, so far, the only assumption on the nonlinear potential is its quadratic form in the field, thus the result is completely general with respect to any kind of quadratic nonlinearity.
\section{Schwarzschild metric}\label{pg:metric}
\begin{figure*}
\includegraphics[width=\linewidth]{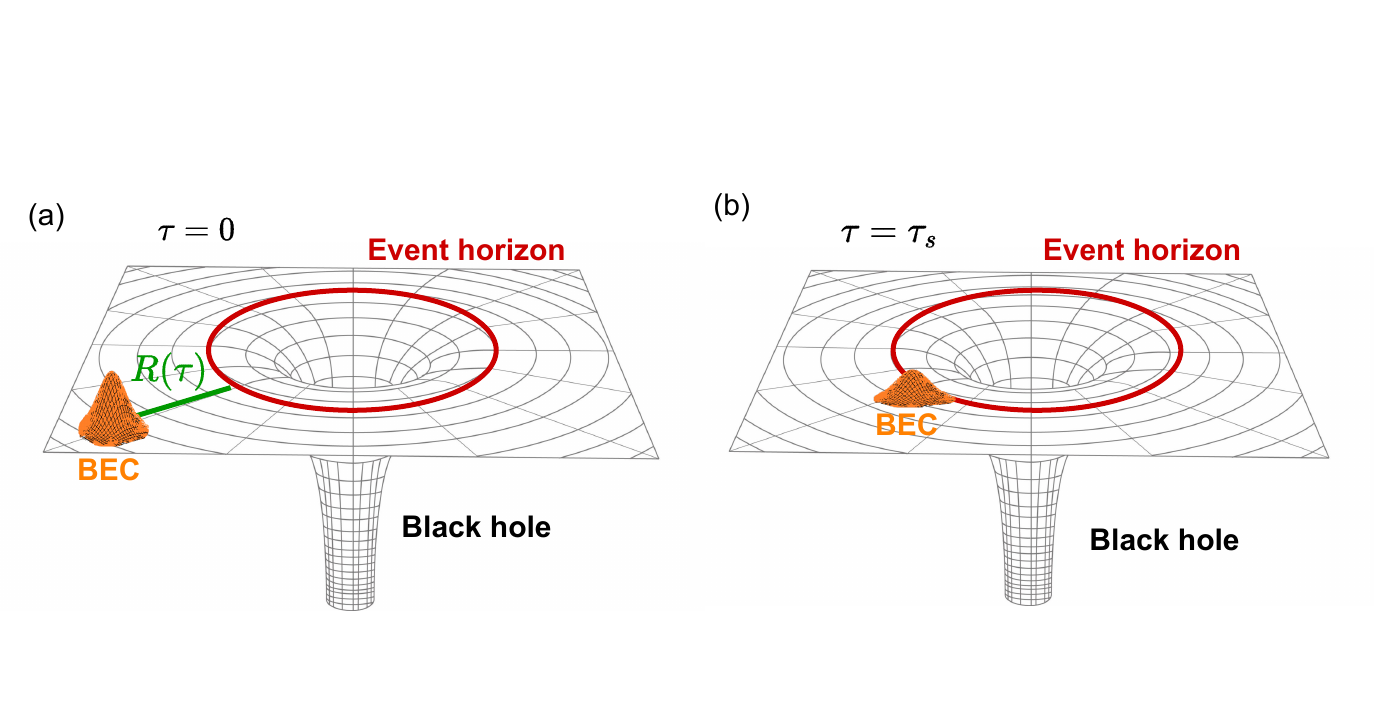}
\caption{\label{fig:illustration} Schematic illustration of the studied phenomenon. A dark matter halo in the form of a self-gravitating Bose-Einstein condensate (BEC) is radially falling toward a Schwarzschild black hole. a) At the beginning of the evolution, the BEC is localized due to self-gravitational interaction. b) At the end of the evolution, the BEC is deformed and, at a critical proper time, destabilizes.
}
\end{figure*}

In the Schwarzschild metric generated by a non-rotating massive object like a static BH
\begin{multline}
ds^2=-X\,dT^2+X^{-1}\,dR^2+R^2\,d\Theta^2\\
+R^2\,\sin^2\Theta\, d\varphi^2
\label{Sch_metric}
\end{multline}
where $T$, $R$, $\Theta$ and $\varphi$ are Schwarzschild coordinates and $X=\left(1-\frac{2GM}{c^2 R}\right)=\left(1-\frac{R_s}{R}\right)$, with $G$ universal gravitational constant, $c$ light speed, $M$ the BH mass and $R_s=\frac{2GM}{c^2}$ Schwarzschild radius.\\
The scenario is sketched in Fig.~\ref{fig:illustration}: the self-gravitating BEC is radially falling toward the BH. We are considering a small DM halo with respect to the BH, thus the spacetime curvature will be the one generated by the BH. The hypothesis of small DM halo is supported by recent studies about space extension of DM halos \cite{wang2020universal}.\\
Eq.~(\ref{general_model_eq}) describes the BEC time evolution, where the effective potential is evaluated using the Schwarzschild metric and its expansion in Fermi normal coordinates. By the metric element expansion in (\ref{g_espansion_elements}) and the Riemann tensor in the Schwarzschild metric, we find the effective potential
\begin{equation}
\begin{split}
V_{\text{eff}}&=\frac{mc^2}{2}g^{(1)00}=\frac{mc^2}{2}R_{0l0m}\, x^lx^m\\
&=\frac{mc^2}{2}(R_{0101}\,x^2+R_{0202}\,y^2+R_{0303}\,z^2)\\
&=\frac{mc^2R_s}{4 R^3(\tau)}(-2x^2+y^2+z^2).
\label{v_eff}
\end{split}
\end{equation} 
We notice that $V_{\text{eff}}$ is time-dependent through $R(\tau)$, which is the radial distance between the BEC and the BH, expressed as a function of the BEC proper time $\tau$. $R(\tau)$ is the solution of the geodesic equation
\begin{equation}
    \begin{split}
        \dot{R}(\tau)=-c\sqrt{E^2-V(R)}
    \end{split}
\end{equation}
where $E$ is a constant and  $V(R)=\left(1+\frac{l^2}{R^2}\right)\left(\frac{R_s}{R(\tau)}\right)$. We verify for this particular choice the validity of the approximation $\dot{R}(\tau)\ll R(\tau)$.\\Moreover the potential (\ref{v_eff}) is quadratic, being attractive in the directions $y$ and $z$ and repulsive along $x$. This difference in sign expresses the presence of tidal forces exerted by the BH on the DM halo. 

Having found the explicit form of the effective potential generated by the Schwarzschild metric, in order to make the problem analytically tractable, we consider the self-gravitating potential.

\section{The highly nonlocal approximation} \label{ASA}
\begin{figure}
\includegraphics[width=\textwidth]{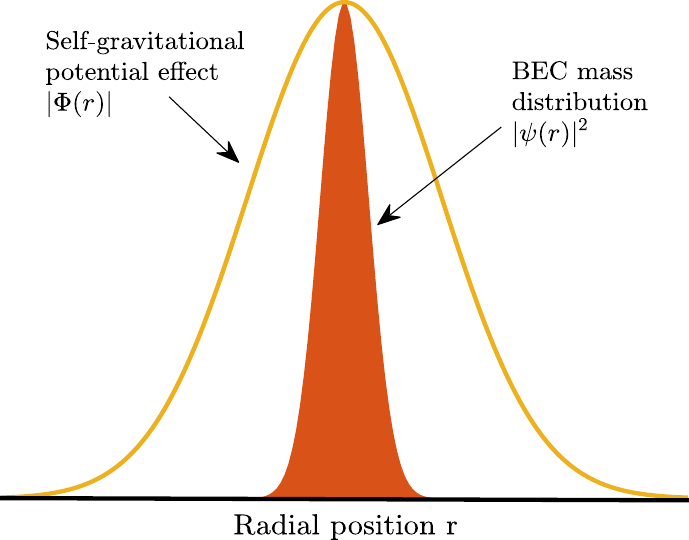}
\caption{\label{fig:nonlocality} Comparison between the spatial density distribution of the BEC $|\psi(r)|^2$ and the intensity of the self-gravitational potential $|\Phi(r)|$. We highlight the nonlocality of the interaction, which spreads beyond the BEC mass distribution.}
\end{figure}

The nonlocality of the self-gravitating potential $\Phi[|\psi|^2]$ is the key property we exploit.
We assume the self-gravitating BEC self-localized; the nonlocal effect of the self-gravitating potential is much wider than the BEC mass distribution (Fig.~\ref{fig:nonlocality}). We hence replace the nonlocal nonlinear potential with a parabolic one. This approach is the so called highly nonlocal approximation (HNA)~\cite{accessible_approx}. The potential $\Phi[|\psi|^2]$ is given by the convolution of the mass density with the Green function
\begin{gather}
\Phi (\vec{r})= \int G(\vec{r}-\vec{r}\,') |\psi (t,\vec{r}\,')|^2 d^3r'; \\
G(\vec{r})=-\frac{G m}{|\vec{r}|} e^{-K|\vec{r}|}.
\end{gather}
The HNA reduces the nonlinear long-range potential to a parabolic one
\begin{equation}
    \Phi\simeq\Phi_M+\tilde{\Phi} r^2=\Phi_M+\frac{\Omega^2}{2} r^2.
\label{phi_parabolic}
\end{equation}
Following the variational method presented in~\cite{accessible_approx}, we find the form for the frequency of the harmonic approximation
\begin{equation}
     \Omega^2=\frac{(4 G)^4}{(3\pi^{1/2})^4\hbar^6}m^{10}N^4.
\end{equation}
We underline that the potential $\Phi$ depends on the BEC total number of particles $N$. \\ Within an additive constant potential term, equation~(\ref{general_model_eq}) becomes
\begin{equation}
\begin{split}
    i \hbar \partial_\tau \psi(\tau,\vec{r})&=-\frac{\hbar^2}{2m}\nabla^2 \psi(\tau,\vec{r})+\frac{m\Omega^2}{2} r^2 \, \psi(\tau,\vec{r})\\
    &+\frac{mc^2R_s}{4 R^3(\tau)}(-2x^2+y^2+z^2) \psi(\tau,\vec{r}).
    \end{split}
    \label{eq_parabolica}
\end{equation}
Equation~(\ref{eq_parabolica}) is characterized by two parabolic terms of potential energy causing the deformation of the BEC DM halo.

\section{Time evolution of the DM condensate}\label{eq_study}
\begin{figure*}
\includegraphics[width=\linewidth]{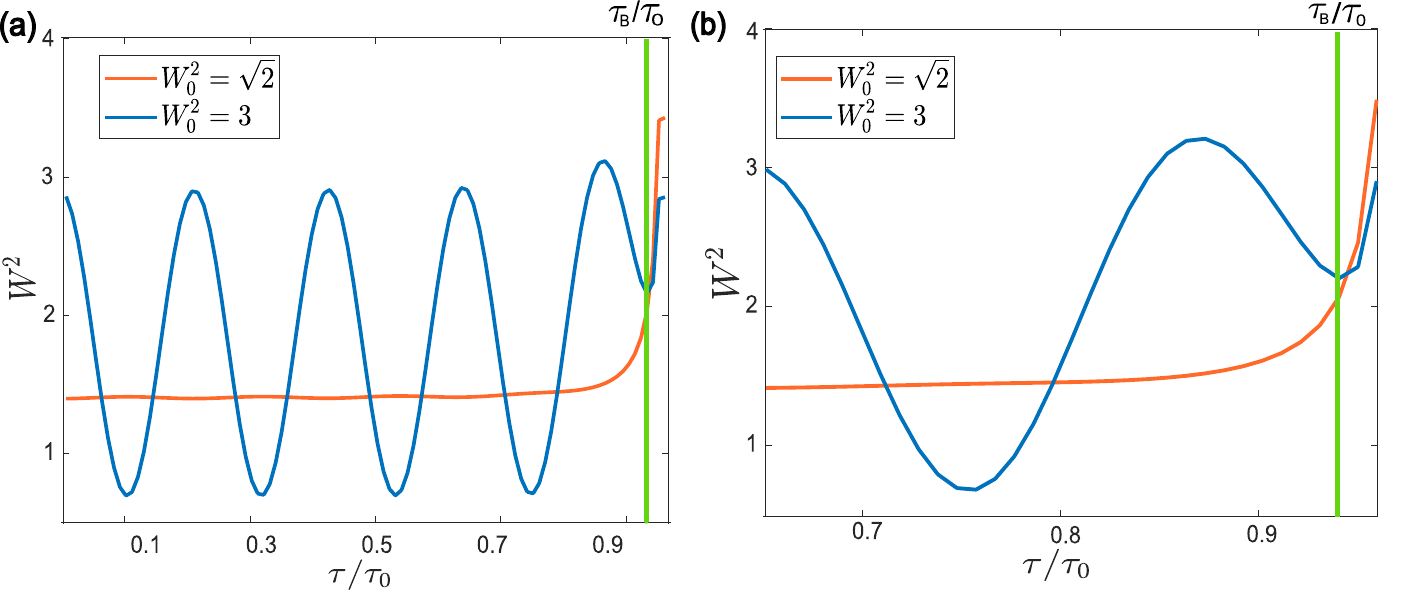}
\caption{\label{fig:width_vs_tau} a) BEC wave function width along the $\xi$ direction during the evolution. Before the breaking time, the mass density distribution has a constant width or an oscillating one according to its initial value. The total potential is attractive and the BEC is in a bound state. After the breaking time, the width follows an exponential trend independently from the previous history. Since the potential is repulsive, the BEC delocalizes. b) Detail of the unstable evolution.}
\end{figure*}

\begin{figure} 
\includegraphics[width=\linewidth]{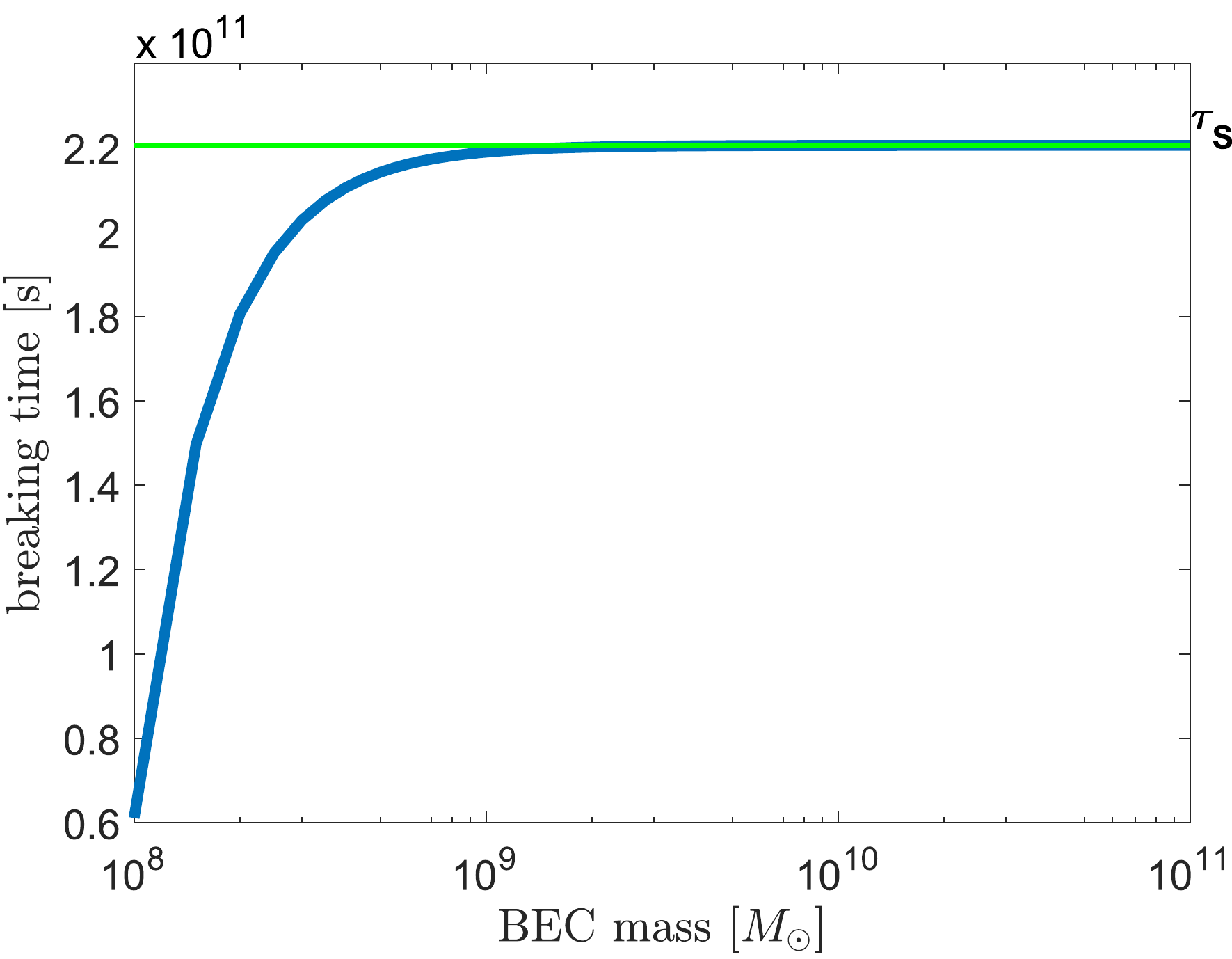}
\caption{\label{fig:tau_B_vs_M} Breaking time versus the BEC mass and Schwarzschild time (green line). }
\end{figure}

A critical element of the effective potential is its time dependence. In our simplified model, it is due to the variation of the radial distance $R(\tau)$, given by the solution of the geodesic equation\\
\begin{equation}
    \begin{split}
        \dot{R}(\tau)&=-c\sqrt{E^2-V(R)}\\
        &=-c\sqrt{\gamma^2-1+\frac{R_s}{R(\tau)}},      
    \end{split}
\end{equation}
where $E$ is a constant equal to $\gamma=1/\sqrt{1-(v/c)^2}$, and assuming the BEC has constant initial velocity $v$, with $\dot{R}(0)=-\gamma v$ for $R(0)\rightarrow\infty$. We also have $V(R)=\left(1+\frac{l^2}{R^2}\right)\left(\frac{R_s}{R(\tau)}\right)=1-\frac{R_s}{R(\tau)}$ for a non-rotating geometry.\\
Moreover, if we take zero initial velocity $v=0$, the solution of the geodesic equation is ($\gamma=1$)
\begin{equation}
    \begin{split}
        R(\tau)=\left(-\frac{3}{2}\sqrt{R_s}c\tau+R^{3/2}(0)\right)^{2/3}.
    \end{split}
    \label{R_tau}
\end{equation}
From Eq.~(\ref{R_tau}), we derive $3$ different time scales: (I) the time when the BEC reaches the position $R(\tau_0)=0$
\begin{equation}
     \tau_0=\frac{2R_0}{3c} \sqrt{\frac{R_0}{R_s}};
\end{equation}
(II) the time when the BEC crosses the event horizon $R(\tau_s)=R_s$, i.e. {\it the Schwarzschild time}
\begin{equation}
    \tau_s=\tau_0\left[1-\left(\frac{R_s}{R_0}\right)^{3/2}\right];
\end{equation}
and (III) the time at which the effective potential dominates over the self-gravitation, i.e. {\it the breaking time}
\begin{equation}
    \tau_B=\tau_0-\frac{2}{3\Omega}.
    \label{tau_B}
\end{equation}
When $\tau=\tau_B$ the effective potential along $x$ becomes equal in strength to the self-gravitating one
\begin{equation}
    \begin{split}
        \frac{1}{2}m\Omega^2 x^2 -\frac{mR_s c^2 }{2R^3(\tau_B)}x^2=0.
    \end{split}
\end{equation}
As a consequence, the sign of the total potential along $x$ changes, and the BEC experiences a repulsion in the $x$-direction.\\
To analytically study the evolution, we consider the dimensionless version of our model equation~(\ref{eq_parabolica})
\begin{equation}
    \begin{split}
        i\, \partial_{\zeta} u(\zeta; \xi, \eta,\theta)&=-\nabla^2 u(\zeta;\xi,\eta,\theta)\\
        &+\frac{1}{2} [\nu^2 - V_\xi(\zeta)\, \xi^2+V_\eta(\zeta)\, \eta^2\\
        &+V_\theta(\zeta)\, \theta^2]\,u(\zeta;\xi,\eta,\theta)\;,
    \end{split}
    \label{eq_adim_3D}
\end{equation}
where we introduce the dimensionless wave function $u=\psi \sqrt{x_p^3}$ and variables
\begin{equation}
    \begin{split}
           \zeta=\frac{\tau}{\tau_p};\, \,\,\, \xi=\frac{x}{x_p}; \, \,\,\, \eta=\frac{y}{x_p}; \, \,\,\, \theta=\frac{z}{x_p};
    \end{split}
\end{equation}
with $\nu^2=\xi^2+\eta^2+\theta^2$, defined by the constants
\begin{equation}
    x_p^2=\frac{\hbar}{\sqrt{2} m\Omega}; \qquad \tau_p=\frac{\sqrt{2}}{\Omega}.
\end{equation}
The dimensionless effective potential terms are
\begin{equation*}
\begin{split}
  &  V_\xi=-\frac{2}{9}\frac{1}{(\zeta-\zeta_0)^2}; \\
    &V_{\eta,\theta}=\frac{1}{9}\frac{1}{(\zeta-\zeta_0)^2}.
    \end{split}
\end{equation*}
As the BEC Hamiltonian is the sum of spatially independent terms, a generic solution is written as the product of three wave functions
\begin{equation*}
    \begin{split}
        u(\zeta, \xi,\eta,\theta)= u_\xi(\zeta,\xi)\, u_\eta(\zeta,\eta)\, u_\theta(\zeta,\theta).
    \end{split}
\end{equation*}
The motion equation splits into three different equations
\begin{equation}
    \begin{split}
      &  i \partial_\zeta u_\xi =-\partial^2_\xi u_\xi +\frac{1}{2} \xi^2 \left[ 1-\frac{2}{9(\zeta-\zeta_0)^2}\right]u_\xi;\\
       & i \partial_\zeta u_\eta =-\partial^2_\eta u_\eta +\frac{1}{2}\eta^2 \left[1+\frac{1}{9(\zeta-\zeta_0)^2}\right]u_\eta;\\
        & i \partial_\zeta u_\theta =-\partial^2_\theta u_\theta +\frac{1}{2}\theta^2 \left[1+\frac{1}{9(\zeta-\zeta_0)^2}\right]u_\theta.
    \end{split}
    \label{eq_adim_3}
\end{equation}
Eqs.~(\ref{eq_adim_3}) have a Gaussian solution
\begin{equation*}
    \begin{split}
        &u_j(\zeta, j)=\frac{1}{\pi^{1/4}\sqrt{q_j(\zeta)}}\, \exp\left[\frac{i}{4}\frac{\dot{q}_j(\zeta)}{q_j(\zeta)}j^2+iP_j(\zeta)\right]\\
        &j=\xi,\eta,\theta.
    \end{split}
\end{equation*}
where $q_j(\zeta)$ and $P_j(\zeta)$ are complex functions: $\frac{\dot{q}(\zeta)}{q(\zeta)}=\frac{1}{R(\zeta)}+\frac{2i}{W^2(\zeta)}$, where $R(\zeta)$ is the curvature radius and $W(\zeta)$ is the width of the wave function. The functions $q_j(\zeta)$ are given by the equations
\begin{equation}
    \begin{split}
\ddot{q}_j(\zeta)+\omega^2_j(\zeta) q_j(\zeta);
\label{eq_q_3}
\end{split}
\end{equation}
with different frequencies in the spatial directions
\begin{equation}
\begin{split}
 &\omega^2_\xi(\zeta)=2\left[1-\frac{2}{9(\zeta-\zeta_0)^2}\right];\\ 
 &\omega^2_\theta(\zeta)=2\left[1+\frac{1}{9(\zeta-\zeta_0)^2}\right]; \\ &\omega^2_\eta(\zeta)=2\left[1+\frac{1}{9(\zeta-\zeta_0)^2}\right].
    \end{split}
    \label{frequenze_tutte}
\end{equation}
Thus, the width in each direction evolves oscillating with a $\zeta$-dependent frequency. In the unperturbed case, the frequency is constant and the equations have oscillating solutions and a stationary one for the initial waist $W_0^2=\sqrt{2}$.\\ Considering specifically the $\xi$-direction, the frequency becomes imaginary, as $\omega_\xi^2<0$ when $\zeta>\zeta_B$, with
\begin{equation}
\begin{split}
 \zeta_B=\frac{\tau_B}{\tau_p}=\zeta_0-\frac{\sqrt{2}}{3}.
    \end{split}
    \label{zeta_b}
\end{equation}
Correspondingly, the solution of the width equation $q_\xi(\zeta)$ from oscillating becomes exponential. As a consequence, the wave function $u_\xi$ ceases to be a Gaussian, the BEC is no more localized and a phase transition occurs.  \\
We show in Figure \ref{fig:width_vs_tau} the width of the $\xi$-dependent solution during the evolution, including the critical case when the initial waist corresponds to the stationary solution of the unperturbed Eq.~(\ref{eq_q_3}). $W$ diverges for times greater than the breaking time $\tau_B$.\\ We consider specifically the dependence of the breaking time $\tau_B$ on the BEC mass (Fig.~\ref{fig:tau_B_vs_M}). Given the BH mass, which fixes the Schwarzschild time $\tau_s$, the breaking time increases with the BEC mass. Below a BEC critical mass, the breaking time is smaller than the Schwarzschild time, thus the DM halo becomes unstable before crossing the event horizon giving rise to an observable instability. we remark that at a time comparable with the breaking time, the perturbative approach is expected not to provide an accurate description of the condensate, and also the mean field approximation is expected to fail, while opening the road to novel and unexplored quantum-gravitational effects.


\section{The astrophysical scenario}\label{astro_exp}
In the astrophysical context, the relevant parameters are the masses: the single particle $m$, the DM halo $M$, and the BH $M_{BH}$, chosen as follows
\begin{equation}
    \begin{split}
     &  \text{BEC:} \quad  m\sim 10^{-22} \text{eV}; \qquad M\sim 10^{6}M_\odot; \\
      & \text{BH:} \quad M_{BH}\sim 10^{9}M_\odot;
    \end{split}
\end{equation}
where $M_\odot$ is the solar mass.\\
The ultralight single particle provides the formation of galactic BECs, in accordance with previous work~\cite{1994, cosmicstructure}. The supermassive BH suits the description of galactic core phenomena.  \\
We have the characteristic frequency of the self-gravitational potential, $\Omega$
\begin{equation*}
    \Omega= \frac{16 G^2 m^5 N^2}{9 \pi \hbar^3}\sim 10^{-11}\, \text{rad/s},
\end{equation*}
and the width of the BEC wave function
\begin{equation*}
    w_0=\sqrt{\frac{\hbar}{m\Omega}}\sim 1.2\, \times 10^{7} \text{m},
\end{equation*}
extending over galactic distances.\\
From the BH mass, we find the Schwarzschild radius
\begin{equation*}
    R_s=\frac{2GM_{BH}}{c^2}\sim 10^{12}\, \text{m}.
\end{equation*}
Assuming a starting distance between the center of the BEC and the BH of order $R_0 \sim  10^5\, R_s \sim10^{17} \text{m}$, we determine the relevant time scales
\begin{equation*}
    \begin{split}
       &\tau_0=\frac{2R_0}{3c} \sqrt{\frac{R_0}{R_s}}\sim 2 \times 10^{11}\, \text{s}\sim 6600 \text{y};\\
      &\tau_s=\tau_0\left(1-\frac{R_s^{3/2}}{R_0^{3/2}}\right) \sim  2 \times 10^{11}\, \text{s}\sim 6600 \text{y};\\
       &\tau_B=\tau_0-\frac{2}{3\Omega}\sim 1.7 \times 10^{11}\, \text{s} \sim 5000\, \text{y}.
    \end{split}
\end{equation*}
DM instability occurs when the breaking time is shorter than the Schwarzschild time, in a way dependent on the BEC mass and the interacting massive objects. The onset of instability may explain some astrophysical observations, including the DM absence in galaxies of different sizes.
\section{The laboratory analog}\label{lab_exp}
\begin{figure} 
\includegraphics[width=\linewidth]{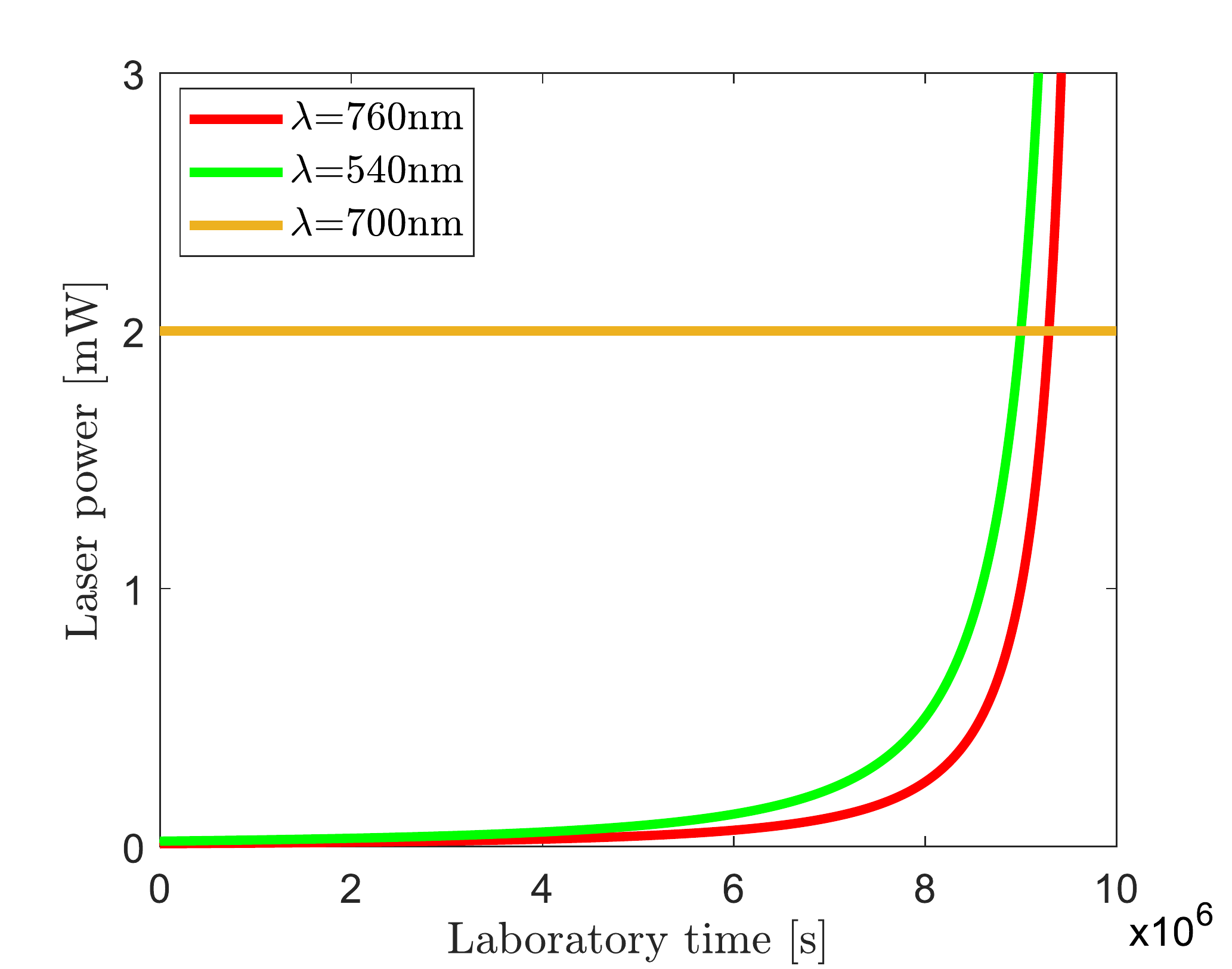}
\caption{\label{fig:laser_power} Laser power of the optical trap of Sodium atoms to mimic the total potential exerted by a Schwarzschild black hole on a galactic BEC. The red-detuned laser beams at $\lambda=700\,\text{nm}$ generate the constant term of the potential, while the time-varying beams mimic the potential felt by the halo during the radial fall. The beam at $\lambda=760\,\text{nm}$ produces an attractive potential while the green laser at $\lambda=540\,\text{nm}$ induces a repulsive potential. }
\end{figure}

\begin{figure*} 
\includegraphics[width=\linewidth]{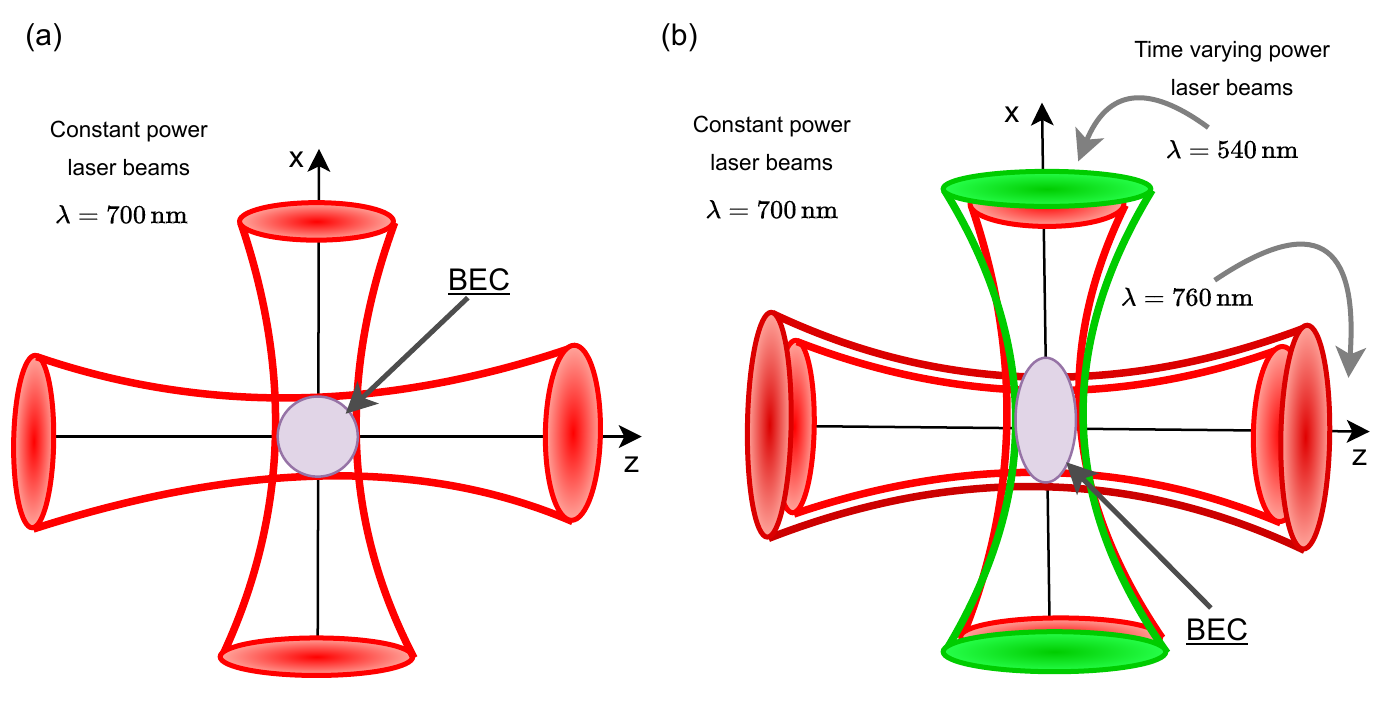}
\caption{\label{fig:optical_trap} Sketch of the optical trap to mimic the gravitational interactions in the astrophysical DM BEC. a) An optical trap with lasers of constant intensity realizes the self-gravitational potential; b) two additional time-dependent laser beams generate the time-varying potential of the BH on the approaching BEC. The detuning between the lasers and the BEC atoms induces deforming tidal forces. The red laser realizes an attractive potential, while the green one is repulsive.}
\end{figure*}

The laboratory realization of an analog system for our model equation~(\ref{general_model_eq}) is based on the production of a BEC from a gas of bosons like Sodium atoms. The condensation is achieved by cooling and confining a Bose gas with tools like a magneto-optic trap \cite{optical_trap_review, laser_cooling_and_trapping}. \\
By using the explicit form for the radius $R(\tau)$ in (\ref{R_tau}), we find that the total potential Eq.~(\ref{v_eff}) is
\begin{equation}
    \frac{m\Omega^2}{2}\vec{r}^{\,2}+\frac{m}{9(\tau-\tau_0)^2}(-2x^2+y^2+z^2).
    \label{potential}
\end{equation}
This potential can be realized by an optical trap with a monochromatic beam of frequency $\omega$ and intensity $I(\vec{r})$ that generates a spatial dependent potential of the form
\begin{equation}
    V(\Vec{r})=\frac{3 \pi c^2 \Gamma}{2\omega_0^3 \Delta }I(\Vec{r})
\end{equation}
where $\omega_0$ is the characteristic resonant frequency of the trapped gas, $\Delta=\omega -\omega_0$ is the laser detuning, and $\Gamma$ is the damping rate due to the radiative energy loss. \\ The constant term of the potential (\ref{potential}) is obtained by a $3$-dimensional optical trap, where $3$ orthogonal laser beams have constant intensity and negative detuning $(\Delta<0)$, namely red-detuned with respect to the condensate characteristic frequency. For the time-dependent term of the potential, we need time-varying intensity beams and a repulsive trap along the $x$-direction. With optical traps, a repulsive potential is obtained by a blue-detuned $(\Delta>0)$ laser beam~\cite{optical_trap_review}. The whole potential is a composite optical trap (Figure~\ref{fig:optical_trap}). The red-detuned laser beams at $\lambda=700\,\text{nm}$ comprise the constant parabolic term analog to the self-gravitation, while the red laser at $\lambda=760\,\text{nm}$ and the green one at $\lambda=540\,\text{nm}$ are the analogs of the time-dependent potential generated by the BH and felt by the BEC during its radial fall. The laser beam intensities vary in time as reported in Figure~\ref{fig:laser_power}.\\
For a BEC of Sodium atoms
\begin{equation*}
    \begin{split}
       & \text{Atomic mass:} \quad  m_A\sim 23\,\text{u.m.a.} \sim 10^{-26}\, \text{kg};\\
        &\text{Resonant frequency:} \quad  \omega_0 \sim 3\times 10^{15}\, \text{rad/s}. \\
    \end{split}
\end{equation*}
Red-detuned laser beams with constant power generate the potential well mimicking the BEC DM self-gravitational potential
\begin{equation*}
    \begin{split}
         \text{Laser wavelength:} \quad &    \lambda \sim 700 \, \text{nm}; \\
        \text{Laser beam waist:} \quad & w_{L}\sim 1 \, \text{mm};\\
        \text{Laser power:} \quad & P\sim 20\, \text{mW};\\
         \text{Detuning:}\quad & \Delta\sim - 3 \times 10^{14} \text{rad/s}.
    \end{split}
\end{equation*}
In the laboratory simulation, the final value $\tau_0$ corresponds to the duration of the experiment. A reasonable value for according to BEC lifetime is $\tau_0\sim 1 \,\text{s}$. To be consistent with our model we assume the DM particle to have mass $m\sim 10^{-22} \,\text{eV}$. With these parameters we can mimic the following astrophysical scenario:
\begin{equation*}
    \begin{split}
        \text{BEC DM waist:}\quad & w_0 \sim 10^{9} \text{m};\\
        \text{BEC total mass:}\quad & M_{\text{BEC}} \sim  10^{4}\, M_\odot;\\
       \text{BH mass:}\quad & M_{\text{BH}}\sim 10^7 M_\odot;\\
       \text{Schwarzschild radius:}\quad & R_s\sim 10^{10}\, \text{m};\\
        \text{Breaking time:}\quad & \tau_B\sim 10^{9} \, \text{s}.
    \end{split}
\end{equation*}
The corresponding breaking time in the laboratory is $\tau_L\sim 0.2\, \text{s}$. 
At this temporal scale, the BEC becomes unstable, and one observes the wave breaking caused by the interaction with the BH. 

\section{Conclusions}\label{conclusions}
Following the hypothesis of the solitonic bosonic nature of interacting dark matter, we investigated the evolution of a Bose-Einstein condensate in curved spacetime. We considered a self-gravitating dark matter halo condensate in the Fermi normal coordinates, which express the d'Alembert operator as the flat Laplacian with parabolic terms generated by the curvature. By a multiple scale expansion, we studied the gravitational interaction with a black hole within highly nonlocal approximation. The resulting effective potential in the GPE is parabolic with a time-dependent frequency.\\ As observed in the literature~\cite{1994}, the collective bosonic wave function explains many of the DM anomalous observations only if the BEC is in its excited states.  In our model, the tidal forces in the effective potential induce exponential deformations and instability in the dark matter halo, which excite higher-order internal modes. The BEC, initially in the ground state of the attractive self-gravitating potential, undergoes a perturbation due to the curved spacetime and evolves into a superposition of excited states.\\ In the direction of the repulsive effective potential, the BEC undergoes a transition from the stable harmonic oscillator to an unstable inverted oscillator~\cite{marcucci2016irreversible}. The instability is reached at a breaking time, which grows with the DM and decreases with the BH mass.  This may be used to explain the presence, or the absence, of DM halos in galaxies of different sizes.
\\ 
Our model is not only a simple closed-form analysis of the interaction of DM with BH but suggests simulating the physics of DM interaction with a curved spacetime in the laboratory by time-varying traps. This may open the way to study first and second quantized dark matter dynamics~\cite{howl2021non, conti2023random} in simulated gravity with atoms, polaritons, and optical systems~\cite{carusotto2013quantum, calvanese_conti_photon_BEC, photon_BEC}.
\begin{acknowledgements}
    We thank prof.~Paolo Maria Santini for the fruitful discussions. 
\end{acknowledgements}

\end{document}